\documentclass[aps,twocolumn,showpacs]{revtex4}

\usepackage{amssymb,amsmath,bm,graphicx,color,subfigure}
\begin{document}
\title{L\'{e}vy targeting  and the principle of  detailed balance}
\author{Piotr Garbaczewski  and  Vladimir Stephanovich}
\affiliation{Institute of Physics, University of Opole, 45-052 Opole, Poland}
\date{\today }
\begin{abstract}
We investigate confined L\'{e}vy flights under  premises of  the principle of detailed balance.
The master equation admits a transformation to L\'{e}vy - Schr\"{o}dinger semigroup  dynamics (akin
to a mapping of the Fokker-Planck equation into the generalized diffusion equation). We solve a stochastic targeting problem for arbitrary
stability index $0<\mu <2$ of L\'{e}vy  drivers:  given  an invariant probability
density function (pdf),  specify the  jump - type  dynamics for which this pdf is a long-time asymptotic target.
 Our (''$\mu$-targeting")  method is exemplified by Cauchy family and  Gaussian target pdfs.
  We solve the reverse engineering problem for so-called  L\'{e}vy oscillators: given a quadratic semigroup potential, find
  an  asymptotic pdf for the associated master equation for arbitrary  $\mu$.

  \end{abstract}
 \pacs{05.40.Jc, 02.50.Ey, 05.20.-y, 05.10.Gg}
\maketitle
\section{Introduction}

Many complex physical systems  (like-wise  non-physical, e.g. economic \cite{stanley}) can be satisfactorily  described   in terms of
 the dynamics of a certain fictitious particle under the action of random forces (noise), originating from   its environment.
 Whenever we can identify a Gaussian noise as an  emergent property of the environment-particle coupling, the  interrelated  notions
 of (thermal) equilibrium,  Boltzmann  asymptotic probability density functions (pdfs) and detailed balance generically follow.
 That is the case in  the standard  Brownian motion picture, based upon kinetic theory derivations, in the presence of (conservative) external forces.

However, in many stochastic systems the experimental data  show that the description based on the introduction of the  Gaussian noise
 is insufficient, since the involved  fluctuations  turn out to generate  have  heavy-tailed distributions of
L\'{e}vy - stable type. Those  distributions are widespread, in  a broad range of systems of varied levels of complexity:  physical,
 chemical, biological \cite{levy1, levy2}, geophysical,  economic  \cite{stanley}.
 That is why a  deeper understanding of properties of  general  complex systems with non-Gaussian noises  is extremely desirable.

For example, contrary to the case of systems with Gaussian fluctuations, in the context of L\'{e}vy flights the notion of "equilibrium",
 although natural under confining conditions,  has no obvious  thermal connotation, see however \cite{stef}.
 It is clear that  any conceivable "thermal equilibrium" concept for non-Gaussian jump-type processes needs to be addressed with  care and should
account for a number of precautions. In particular, an issue of physically motivated  thermalization mechanisms  for (confined) L\'{e}vy flights
 has received  only a residual  attention in the literature, \cite{tsallis,stariolo,borland} and \cite{stef,stef1}. The  main obstacle here may
  be that   the  source of L\'{e}vy noise is interpreted  as extrinsic to the
physical system under consideration,  with no reliable kinetic theory background, i.e. with no identifiable
microscopic channels of an energy exchange with the environment.

L\'{e}vy flights  are pure jump (jump-type) processes. Therefore, it seems useful  to recall  that various
model realizations of  standard   jump processes (jump size is bounded from below and above) can be thermalized,
   by means of a  locally defined scenario of an energy exchange with the thermostat, \cite{maes,maes1,maes2} see also  \cite{burda}.
    It amounts to a suitable re-definition of transition rates for the  jump  process which enforces
    the {\em  {principle of detailed balance}}  to be respected by a random motion.
    We shall elaborate upon extension of this idea  to  L\'{e}vy - stable processes, with a focus on the existence of  asymptotic (large time limit) invariant pdfs, of the manifest Boltzmann  form.

Our approach is close to that used to analyze  the L\'{e}vy motion in  systems with topological complexity like polymers (see, e.g. \cite{sokolov}),
 but   remains distinctively different from a  standard theory of confined
L\'{e}vy flights  which is based on the Langevin modeling.
There is no known (additive or multiplicative) Langevin representation for   L\'{e}vy processes respecting the  canonical form of
detailed balance.

 The considered   class of confined L\'{e}vy flights  is   well  suited for the description of jump-type processes that are
equilibrated (eventually, to a thermal equilibrium state)  by a mild   spatial  disorder    of the physical  environment in which jumps take place.
  The inhomogeneity of the environment  is quantified by turning over from the master equation to  the  affiliated  semigroup dynamics. It is a
 suitable functional form of the   semigroup potential (which we consider to be a continuous function) that allows for  a unique  asymptotic
    invariant state.  That ensures the existence  of an  asymptotic invariant pdf  for the master  equation in question.

The structure of the paper is as follows.  First we discuss an issue of detailed balance for standard jump processes and next define its immediate generalization to L\'{e}vy flights ($\mu $-family of L\'{e}vy-stable laws with $0<\mu \leq 2$), Sections II and III. A mapping of  the resultant  master equation to a fractional version of the generalized diffusion equation follows in Section IV. For clarity  of presentation, we make a Brownian detour in Section V to indicate how the  semigroup framework is  related  to the standard  Fokker-Planck dynamics  of diffusion-type processes. In Section VI we describe the L\'{e}vy  $\mu$ - targeting under an assumption that target pdfs are selected from so-called Cauchy  $\alpha$ - family  of pdfs.  For a computationally advantageous example of $\alpha =2$ and arbitrary $\mu \in (0,2)$ we  provide analytic formulas for the associated semigroup potentials (they define the semigroup dynamics which makes the considered pdfs to be genuine asymptotic targets of the jump-type process).  In Section VII the L\'{e}vy targeting is considered for Gaussian target pdfs.  Section VIII presents a complete solution of the reverse engineering problem for the $\mu $-family of L\'{e}vy oscillators, corresponding to quadratic semigroup potential. The obtained analytic formulas for asymptotic pdfs are depicted in Figs. 4  and 5. Not to overburden the paper with formal arguments, a general solution of the reverse engineering problem for arbitrary semigroup potential has been moved to another publication.

  \section{Jump processes and  detailed balance}

  Let $K$ be a finite state space, with $x, y \in K$. We consider  Markovian  stochastic dynamics for a finite  random  system,  with transition rates  $k(x|y) \equiv  k(y \rightarrow x)$. Given an initial probability distribution $\rho_0(x)$,
its time evolution for times $t \geq 0$ is governed by the master equation:
 \begin{equation} \label{tx1}
 {\frac{d}{dt}} \rho _t(x) = \sum_{y\in K} [k(x|y)\rho _t(y) -  k(y|x)\rho _t(x)]  \, .
\end{equation}
Given  a stationary solution $\rho _{eq}(x)$  of the master
equation, $\dot{\rho }_{eq}(x)=0$.
 If we  have
\begin{equation} \label{tx2}
k(x|y)\rho _{eq}(y)=  k(y|x)\rho _{eq}(x)
\end{equation}
one says  that the \it  condition of detailed balance  \rm  is
fulfilled.

Let  $\rho_{eq}(x) \propto \exp[-U(x)]$, where $U$ is a   suitable  function on $K$.  (The inverse temperature $\beta $ can be safely absorbed in the definition  of $U$.  As well, for clarity of discussion,  we can set $\beta =1$).
 Accordingly:
\begin{equation} \label{tx3}
k(x|y) = k(y|x) \exp [U(y)-U(x)] \, .
\end{equation}

We note that $k_0(x|y) = k_0(y|x)$, in a finite  state space, yields a uniform distribution $\rho _{eq}(x)= const$ for all $x\in K$. Let us consider  a simple multiplicative   modification of a symmetric transition intensity $k_0(x|y)$:
\begin{equation} \label{k0xy}
k_0(x|y)  \Longrightarrow k_U(x|y) = k_0(x|y) \exp
\left[{\frac{U(y)-U(x)}{2}}\right]
\end{equation}
By inspection  (simply replace $k(x|y)$ by $k_U(x|y)$  in Eqs.~\eqref{tx1}- \eqref{tx3}) one  verifies  the  validity of the  detailed balance condition, with $\rho_{eq}(x) \propto \exp[-U(x)]$ as the corresponding
stationary distribution.

We assume that an equilibrium density $\rho _{eq}(x)>0$ is unique and  presume  the detailed balance condition \eqref{tx2}, \eqref{tx3} to be respected.  Then,  the relative entropy  (negative of the Kullback-Leibler entropy) becomes
\begin{equation}
{\cal{S}}(\rho _t|\rho _{eq})
= \sum_{x\in K} \rho _t(x) \ln {\frac {\rho _t(x)}{\rho _{eq}(x)}} =
{\cal{F}}(\rho _t)-{\cal{F}}(\rho _{eq}) \geq 0.
\end{equation}
Here an obvious analogue of the familiar Helmholtz free  energy  ${\cal{F}}(\rho _t) = \sum_{x\in K} U(x)\rho _t(x) -  {\cal{S}}(\rho _t)$  has been introduced,  with ${\cal{S}}(\rho _t)= - \sum_{x\in K} \rho _t(x) \ln \rho _t(x)
$ being  the Shannon entropy of the probability distribution $\rho
_t(x)$.  We have  ${\cal{F}}(\rho _t)   \geq {\cal{F}}(\rho _{eq})  = -  \ln \sum_{x\in K} \exp[-U(x)]$.
The relative entropy is monotonous in time and converges to zero, which is accompanied by a decrease of the free energy ${\cal{F}}(\rho _t)$ to its minimal value ${\cal{F}}(\rho _{eq})$.

It is useful to mention an interesting inverse stationary problem of Refs. \cite{maes1,maes2}. Namely, for an arbitrary
positive probability distribution $\rho _{eq}(x)>0$ on $K$ there exits  a   function  $U(x)$ such that $\rho _{eq}(x)$ is invariant under the jump dynamics  with the transition rate $k_U(x,y)$ of the form \eqref{k0xy}. In the original formulation of  Ref. \cite{maes2},  the reference transition rate  $k_0(x,y)$ needs not to be symmetric.

\section{Detailed balance for L\'{e}vy flights}

The above reasoning  gives an immediate  justification to the strategy adopted before  in the context of L\'{e}vy -stable processes, albeit  with no explicit reference to the detailed balance principle,  in a number of papers \cite{sokolov, geisel, belik}.  We also note Refs. \cite{stef,stef1,gar},  where  "stochastic targeting" and  related "inverse engineering" (terms, originally coined in Ref. \cite{klafter}) have been exploited to this end.

To proceed further, we recall that a characteristic function of a random variable $X$  completely determines a probability distribution of that variable. If this distribution admits a pdf $\rho(x)$, we can write $<\exp(ipX)> = \int_R \rho (x) \exp(ipx) dx$. A  classification of infinitely divisible probability laws  is provided by the   L\'{e}vy-Khintchine formula for the exponent $- F(p)$ of   $<\exp(ipX)> =  \exp[-F(p)]$.

We restrict subsequent considerations  to a subclass of stable probability distributions with $F(p) =  |p|^{\mu }$,
with   $0< \mu \leq 2$. The induced  jump-type   dynamics $<\exp(ipX_t)>= \exp[-t F(p)]$ is  conventionally interpreted in terms of L\'{e}vy flights  and  quantified  by means of a pseudo-differential (fractional) analog of the heat equation for corresponding pdf
\begin{equation} \label{master}
\partial _t \rho  = -  |\Delta |^{\mu /2} \rho =  \int [w_{\mu }(x|y) \rho (y) - w_{\mu }(y|x) \rho (x)]dy,
\end{equation}
which has been  rewritten as a  master equation for a random system on real axis, with a pure jump
dynamics.  The jump rate  $ w_{\mu } (x|y)\propto 1/|x-y|^{1+\mu }$ is  a symmetric  function,
$w_{\mu }(x|y)= w_{\mu } (y|x)$ akin to $k_0(x|y)$ of the previous subsection. We recall that the action of a fractional operator $|\Delta |^{\mu /2}$ on a function from  its domain is defined by means of the Cauchy
principal value of an involved  integral:
\begin{equation}
 -(|\Delta |^{\mu /2} f)(x)\,  =
   {\frac{\Gamma (\mu +1) \sin(\pi \mu/2)}{\pi }} \int  {\frac{f(z)- f(x)}{|z-x|^{1+\mu }}}\,
 dz \,\, .
\end{equation}

Mimicking the previous step \eqref{k0xy}, we open a possibility of a  locally  controlled energy exchange with an   environment, by  modifying   the jump rate $ w_{\mu}(x|y)$   of the free  (neither external forces nor potentials)
fractional dynamics    to the  non-symmetric  form  $w_{\mu }^U(x|y) \neq w_{\mu }^U(y|x)$:
$w_{\mu}^U(x|y) = w_{\mu }(x|y) \, \exp ([U(y) - U(x)]/2)$.
With $w_{\mu }^U(x|y)$ replacing $w_{\mu }(x|y)$,   the master equation  (\ref{master})  ultimately takes
  a slightly discouraging form, known from a number of previous publications:
  \begin{widetext}
\begin{equation}\label{kinetic}
\partial _t \rho = -  |\Delta |^{\mu /2}_U \rho =    \int [w_{\mu }^U(x|y) \rho (y) - w_{\mu }^U(y|x) \rho (x)] dy =
-  [\exp (-U/2)]\, |\Delta |^{\mu /2}[ \exp(U/2 )
    \rho ]   +    \rho \exp (U/2 ) |\Delta |^{\mu /2} \exp(-U/2)\, .
\end{equation}
\end{widetext}
The  above  transport equation cannot be transformed to any known form of the  fractional Fokker-Planck dynamics, based on the standard (L\'{e}vy-stable) Langevin modeling, (c.f. \cite{fogedby}-\cite{dybiec} for  literature sample). These two dynamical patterns of behavior are inequivalent, \cite{gar,stef}.

For a suitable (to secure normalization)  choice of  $U(x)$, $\rho _{eq}(x) \propto \exp [- U(x)] $ is a stationary solution of Eq.~(\ref{kinetic}). The  detailed balance principle of the form \eqref{tx2}, \eqref{tx3}  holds true.

    For the record, let us mention that  the free  fractional Fokker-Plack equation   (\ref{master}) has  no stationary solutions.  Thus, the jump-type  dynamics  with properly modified jump rates  clearly  may  give rise  to  confined
L\'{e}vy flights. Their asymptotic pdfs  in principle may have an arbitrary, not necessarily finite and/or small,  number of moments. The reference stable laws  generically  have no  moments of order higher than one.

\section{L\'{e}vy semigroup modeling}

The master equation  (\ref{kinetic}) cannot be derived within  the  standard Langevin modeling of confined
L\'{e}vy flights, \cite{gar,stef1,gar1}. The latter motion scenario (with an ample coverage in the literature, \cite{fogedby,chechkin,dubkov}) is incompatible with that based on  the detailed  balance principle \eqref{tx2}, \eqref{tx3} and the resultant  Eq.~(\ref{kinetic}),  c.f. \cite{stef,gar}.

 The form of Eq.~(\ref{kinetic})  is not handy.
However, there exists an equivalent description of the  pertinent dynamics
in terms of a   L\'{e}vy-stable semigroup or a fractional (L\'{e}vy-) Schr\"{o}dinger-type equation,  \cite{geisel,belik,stef1,gar1}. The difference with pure time-dependent Schr\"{o}dinger equation is the absence of imaginary unit $i$ before time derivative (e.g. in Eq. \eqref{kinetic}).

To this end let us consider  the L\'{e}vy-Schr\"{o}dinger Hamiltonian operator  with an  external potential
\begin{equation}\label{hamiltonian}
\hat{H}_{\mu }  \equiv   |\Delta |^{\mu /2} +  {\cal{V}}(x)\, .
\end{equation}
Suitable  properties of ${\cal{V}}$ need to be assumed, so that $-\hat{H}_{\mu }$ is a legitimate  generator  of a dynamical semigroup
  $\exp(-t \hat{H}_{\mu })$ and     $\partial _t \Psi = \hat{H}_{\mu } \Psi$  holds true for real functions $\Psi (x,0)\rightarrow \Psi (x,t)$.

Let us a priori select an invariant probability density  $\rho _{eq}(x) \doteq \rho _*(x)\propto \exp[-U(x)]$  of Eq.~(\ref{kinetic}). To make it an asymptotic pdf of a well defined  jump-type  process we address  an issue of the existence of a suitable semigroup dynamics.

Looking for stationary solutions of the affiliated  semigroup  equation
$\partial _t \Psi = \hat{H}_{\mu } \Psi$, we realize that if a square root of a  positive  invariant   pdf $\rho _*(x)$  is asymptotically  to  come out via  the semigroup dynamics  $\Psi \rightarrow  \rho _*^{1/2}$, then the resulting
fractional Sturm-Liouville equation $\hat{H}_{\mu }  \rho _*^{1/2}=0$ imposes a {\em{compatibility condition}}
 upon the functional form of  ${\cal{V}}(x)$, that needs to be respected. Namely, the potential function and invariant pdf $\rho _*^{1/2}$ should be related as
 \begin{equation}\label{comp}
 {\cal{V}}  =   -\frac{|\Delta |^{\mu /2}  \rho ^{1/2}_*}{\rho ^{1/2}_*}.
 \end{equation}
 The resulting semigroup dynamics provides a solution for  the L\'{e}vy stable {\em {targeting problem}}, with a predefined invariant pdf.

Inversely,  if we predefine a concrete potential function ${\cal{V}}(x)$, then the functional form of an asymptotic  invariant  pdf  $\rho _*(x)$ (actually $\rho _*^{1/2}(x)$) comes out from the  above  compatibility condition. We call the problem of derivation of $\rho _*$ from a predefined semigroup potential ${\cal{V}}(x)$ as {\em  {reverse engineering problem}} , see Ref.~\cite{klafter} where this idea had been put forward.

For ${\cal{V}}={\cal{V}}(x)$ bounded from below,  the  integral kernel $k(y,s,x,t)=\{ \exp[-(t-s)\hat{H}] \} (y,x)$, $s<t$, of the dynamical semigroup $\exp(-t\hat{H})$  is  positive. The semigroup dynamics reads:
$\Psi  (x,t) = \int \Psi (y,s)\, k(y,s,x,t)\,  dy$ so that for all $0\leq s<t$  we can reproduce the dynamical pattern of  behavior,    actually  set by  Eq.~(\ref{kinetic}), but now in terms of Markovian pdfs $p(x,s,y,t):$
\begin{equation} \label{un3}
\rho (x,t) =  \rho _*^{1/2}  (x)\Psi (x,t) = \int p(y,s,x,t) \rho (y,s) dy,
\end{equation}
where  $$p(y,s,x,t) = k(y,s,x,t)\frac{\rho _*^{1/2}(x)}{\rho _*^{1/2}(y)}.$$
An asymptotic behavior of $\Psi (x,t)\to \rho _*^{1/2}(x)$ implies  $\rho (x,t) \rightarrow \rho _*(x)$.

A remark is in place here. The spectral theory of fractional operators of the form (\ref{hamiltonian}) has received a broad coverage in the mathematical  \cite{davies,bertoin,kaleta,lorinczi,lorinczi1} and mathematical physics  literature \cite{carmona,carmona1}.  An explicit  functional form of  asymptotic invariant pdfs  of confined L\'{e}vy flights  $\rho _*$    ($\rho _*^{1/2}$ in the semigroup notations) is seldom accessible, with a  notable exception of those for  Cauchy flights \cite{dubkov,gar}.  Therefore it is wise to rely on accumulated data that are available,
 about  the near-equilibrium behavior and the decay of pdfs as $|x|\to \infty$, under very general circumstances.
  Various rigorous estimates pertaining   to the decay at infinities of the eigenfunctions, quantify  the number of moments of the associated  pdfs for  different  classes of potential functions ${\cal{V}}(x)$.  As well, fractional versions of Feynman-Kac formula determining an integral kernel of the semigroup operator, and thence the transition probability which generates (by virtue of Eq. \eqref{un3}) the pdf $\rho (x,t)$ dynamics consistent with   Eq.~(\ref{kinetic}),  have an ample coverage therein.

\section{Brownian detour}

The aim of this section is to describe the relation between above L\'{e}vy - Schr\"{o}dinger semigroup framework and standard Fokker-Planck dynamics  of diffusion-type processes. To make this description clear, here we put explicit relations, translating things from the language of partial differential equations (like Fokker-Planck one) and dealing explicitly with pdfs into the operator language, inherent in (both normal and fractional) quantum mechanics and ultimately in L\'{e}vy - Schr\"{o}dinger semigroup.

In the theory of  standard Brownian motion, the  Langevin equation or the like (stochastic differential equation with
the Wiener noise input) allows to infer  a corresponding Fokker-Planck equation. This in turn can be transformed into a Hermitian (strictly-speaking, self-adjoint) spectral  problem, \cite{risken}. Contrary to the L\'{e}vy-stable case, for diffusion-type processes  both these  descriptions  (e.g. semigroup and Langevin-based Fokker-Planck approaches) are similar descriptions of  the dynamics of $\rho (x,t)$.

Given the spectral solution for the operator $\hat{H}= - \Delta + {\cal{V}}$,  the  integral kernel of   $\exp(-t\hat{H})$
reads $k(y,x,t)= \sum_j \exp(- \epsilon _j t) \, \Phi _j(y) \Phi ^*_j(x)$. Here, the sum may be replaced by an integral in case of a continuous spectrum and (generalized)  eigenfunctions may   be complex-valued.

If we set ${\cal{V}}(x)=0$ identically, a purely  continuous spectral problem arises. Then,  one arrives at the
familiar heat kernel
\begin{eqnarray*}
&&k(y,x,t)= [\exp(t\Delta )](y,x) = \\
&&(2\pi )^{-1/2} \int \exp(-p^2t)\, \exp(ip(y-x))\, dp= \\
&&(4\pi t)^{-1/2}\, \exp\left[-\frac{(y-x)^2}{4t}\right],
\end{eqnarray*}
which is a well-known transition probability density of the Wiener process (actually, upon setting $t \to (t-s)$).

When confining potentials are present, either entire spectrum or its part turns out to be discrete, the corresponding eigenfunctions being real-valued. A  standard example is the  harmonic oscillator i.e. the Ornstein - Uhlenbeck process in its original  stochastic  version. Consider
\[ \hat{H}= (1/2)(-\Delta + x^2 -1).\]
The integral kernel of $\exp(-t\hat{H})$ is given by the classic Mehler formula \cite{mehler}:
\begin{eqnarray*}
&&k(y,x,t) = k(x,y,t)= \exp(-t\hat{H})(y,x)= \\
&&=\frac{1}{\pi \sqrt{1-e^{-2t}}}\exp\left[-\frac{x^2-y^2}{2} - \frac{(xe^{-t} - y)^2}{1-e^{-2t}}\right].
\end{eqnarray*}
The normalization condition
\[\int k(y,x,t) \exp[(y^2-x^2)/2]\, dy =1\]
directly  employs (and defines upon setting  $t \to (t-s)$) the transition probability density of the Ornstein-Uhlenbeck process,
\[p(y,x,t) = k(y,x,t)\frac{\rho _*^{1/2}(x)}{\rho _*^{1/2}(y)}\]
with $\rho _*(x)=\pi ^{-1/2} \exp(-x^2)$ being its (Gaussian) invariant pdf.

\section{Cauchy family of pdfs  and L\'{e}vy $\mu$ - targeting}

Here we describe in some detail  the L\'{e}vy stable (with stability index $\mu$)
targeting strategy with the pre-determined one-parameter  family of Cauchy target pdfs:
\begin{equation}\label{mutar1}
\rho _*(x) \equiv \rho_\alpha (x)=\frac{\Gamma(\alpha)}{\sqrt{\pi}\Gamma(\alpha -
1/2)}\frac{1}{(1+x^2)^\alpha},\ \alpha>1/2.
\end{equation}
We consider functions \eqref{mutar1} as asymptotic invariant  pdfs for the stochastic jump-type  process of
Eq. ~\eqref{kinetic}. We  wish to demonstrate that any $\mu $-stable driver can be employed to this end.

 Instead of addressing directly Eq.~(\ref{kinetic}), we use the  semigroup dynamics $\exp(-t \hat{H}_{\mu })$  generated by the fractional operator (\ref{hamiltonian}), i.e.  the integro-differential  equation
\begin{equation}\label{mutar2}
  \partial_t \Psi=-|\Delta|^{\mu/2}\Psi -{\cal V}_\mu \Psi,
\end{equation}
where  $\Psi(x,t)\equiv \rho(x,t)/\rho_*^{1/2}(x)$  and ${\cal{V}}_{\mu }(x)= - (|\Delta |^{\mu /2}\rho ^{1/2}_*)/\rho ^{1/2}_*$, $0<\mu \leq 2$.

We note  that the  Cauchy  family \eqref{mutar1} has been chosen for computational  convenience only. In principle, there is no restriction
 on the choice of any other target pdf  $\rho _*(x)$.  The qualitative outcome will be the same as that provided in terms of family (\ref{mutar1}). Hereafter we call such general procedure "$\mu$-targeting".

Let us add,  as a side comment,  that the Cauchy family   of pdfs has played  an important role in the previously mentioned search for "thermodynamic equilibria",  that may possibly be associated with confined  L\'{e}vy flights,  \cite{tsallis,stariolo,borland}.
 It is known  \cite{tsallis,stef},  that an exponent $\alpha $ in principle   can be directly related to the thermal equilibrium
 label $\alpha \propto  1/ k_BT$. An analogous observation has been reported in Refs.~\cite{stef,stef1}, after transforming pdfs  (\ref{mutar1}) into an  "exponential form", which resembles Boltzmann  one  $\rho _*\propto \exp  (-U)$, with $U(x)= \alpha \ln (1+x^2)$.

To pass over to the semigroup description we need to infer  ${\cal V}_\mu(x)$, given $\rho _*$.
This can be done  analytically by means of the Fourier transform, specifically because Fourier images of functions  \eqref{mutar1} for arbitrary $\alpha >0.5$  exist  in a closed analytical form of MacDonald functions $K_\nu$ \cite{abr}.

The Fourier image   $g(k)=\frac{1}{\sqrt{2\pi}}\int_{-\infty}^{\infty}g(x)e^{\imath kx}dx$   of a function $g(x)$, when adopted to  $g(x)=|\Delta|^{\mu/2}f(x)$ reads $|k|^\mu f(k)$.
Fourier images of the square roots of pdfs  \eqref{mutar1} read
\begin{equation}\label{mutar4}
\rho_\alpha^{1/2} (k)= \sqrt{\frac{2\Gamma\left(\frac{1+\alpha}{2}\right)}{\pi \Gamma\left(\alpha -1/2\right) \Gamma(\alpha/2)}}\
|k|^{\frac{\alpha -1}{2}}K_{\frac{\alpha -1}{2}}(|k|).
\end{equation}
An  explicit expression for  the $\alpha $-family of   "$\mu$ - potentials"  ${\cal{V}}_{\mu ,\alpha }(x)= - (|\Delta |^{\mu /2}\rho ^{1/2}_{\alpha })/\rho ^{1/2}_{\alpha }$  readily  follows
\begin{eqnarray}\label{mutar5}
&&{\cal V}_{\mu,\alpha}(x)=-\frac{2^\mu}{\sqrt{\pi}} (1+x^2)^{\alpha/2}\frac{\Gamma \left(\frac{1+\mu}{2}\right)\Gamma \left(\frac{\alpha+\mu}{2}\right)}{\Gamma\left(\frac{\alpha}{2}\right)} \times \nonumber \\
&&\times {_2}F_1\left(\frac{1+\mu}{2}, \frac{\alpha+\mu}{2},\frac 12, -x^2\right),
\end{eqnarray}
where ${_2}F_1(a,b;c,x)$ is a hypergeometric function \cite{abr}.

The expression \eqref{mutar5} gives the general form of the semigroup potentials ${\cal{V}}_{\mu ,\alpha }(x)$ for arbitrary $\alpha$ and $\mu$. To have a better feeling about the properties of the function \eqref{mutar5}, we should
explore this expression for some specific values of parameter $\alpha$. Further discussion is limited to the  case  of  $\alpha=2$, i.e.
\begin{equation}\label{mutar6}
    \rho_2^{1/2}(x)=\sqrt{\frac{2}{\pi}}\frac{1}{1+x^2} \rightarrow   \rho_2^{1/2}(k)=e^{-|k|}.
\end{equation}
We note that the  Fourier image $\rho_2^{1/2}(k)$  directly  comes from the  general expression \eqref{mutar4}, if we  use  $K_{1/2}(x)=(\pi/2 x)^{1/2}\, e^{-x}$. Then  for all $0<\mu <2$ we have
\begin{equation}\label{mutar7}
{\cal V}_{\mu,2}(x)=-(1+x^2)^{\frac{1-\mu}{2}}\Gamma(1+\mu) \cos\left[(1+\mu)\arctan x\right] .
\end{equation}

For $\mu=1$ from \eqref{mutar7} we recover  our elder result, originally obtained in  the context of Cauchy flights, \cite{gar}:
\begin{equation}\label{mutar10}
   {\cal V}_{1,2}(x)= \frac{x^2-1}{1+x^2}.
\end{equation}
The plots of the  $\mu $-dependence of  \eqref{mutar7} are reported on Fig.1.

\begin{figure}
\centerline{\includegraphics[width=1.2\columnwidth]{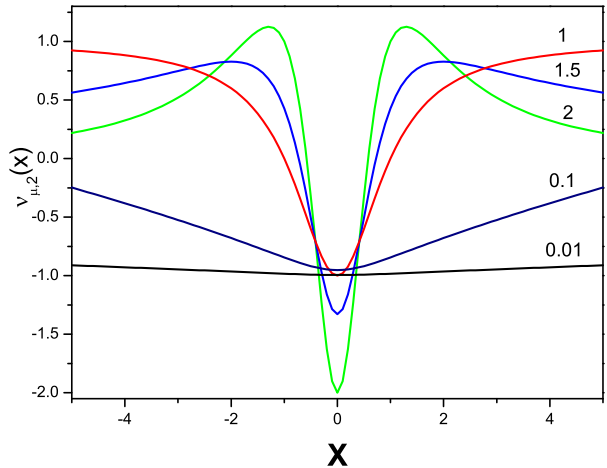}}
\caption{Dependence ${\cal V}_{\mu,2}(x)$ for $\rho_2$ terminating pdf. Figures near curves correspond to $\mu$ values. The potentials for $\mu=1$ and $2$ are given by Eqs. \eqref{mutar10} and \eqref{mutar8} respectively.}
\label{f:pop}
\end{figure}

The stability index $\mu $ is constrained to stay within  an  interval $0< \mu \leq 2$.  The boundary value $\mu =2$ takes us beyond the jump-type  "territory"  to continuous (Wiener noise)  stochastic processes. It is interesting to observe  that on the level of  "$\mu $-potentials", the transition   from $\mu <2 $ to $\mu =2 $ is actually smooth.

Analytically,  recalling the fractional derivative transcription   $(- \Delta )^{\mu /2}$ $\equiv$  $ -  \partial ^{\mu }/\partial |x|^{\mu }$ and   then   setting "blindly"   $\mu=2$  in \eqref{comp}, we   arrive  at    the   semigroup potential for the operator $\hat{H}= - \Delta +  {\cal V}_{2,2}$:
\begin{equation}\label{mutar8}
    {\cal V}_{2,2}(x)={\cal V}_{FP}(x)=\frac{\frac{d^2}{dx^2}\rho_2^{1/2}(x)}{ \rho_2^{1/2}(x)}=\frac{2(3x^2-1)}{(1+x^2)^2}.
\end{equation}
The notation ${\cal V}_{FP}(x)$ refers to the fact that this potential appears  in the semigroup (self-adjoint) version, (c.f. Ref. \cite{risken}) of the standard Fokker-Planck equation for a diffusion-type process.
The same result \eqref{mutar8} can be obtained from Eq. \eqref{mutar7} at $\mu=2$.

 The expression \eqref{mutar7} permits us to
expand the potential ${\cal V}_{\mu,2}(x)$ near $\mu=2$ to obtain
\begin{widetext}
\begin{equation}\label{mutar12}
   {\cal V}_{\mu \to 2,2}(x)\approx \frac{2(3x^2-1)}{(1+x^2)^2} -\frac{\mu-2}{(1+x^2)^2}\left[2x(x^2-3)\arctan x
   +(3x^2-1)\left(2\gamma -3+\ln(1+x^2)\right)\right],
\end{equation}
\end{widetext}
where $\gamma \approx 0.577216$ is Euler constant.
This (along with numerical curves from Fig.1) demonstrates the continuous transition   from $\mu<2$ to  $\mu=2$ in  ${\cal V}_{\mu,2}(x)$.

\section{Gaussian $\mu$-targeting for L\'{e}vy flights}

In the previous publications \cite{stef,stef1,gar} we have investigated various patterns of jump-type  and diffusive behavior that would produce a priori selected, basically heavy-tailed pdfs  in the large time asymptotics. While an association of jump type-processes with pdfs possessing a finite number of moments is rather natural, an observation of Ref.~\cite{stef1} that diffusion-type processes may as well admit such  asymptotic pdfs, may be classified as  "unnatural".

Here we proceed in the very same "unnatural" vein, asking for  a L\'{e}vy-stable jump-type dynamics, whose asymptotic pdf would have a definite Gaussian form.
Let us  select  the Gaussian  target  pdf
\begin{equation}\label{dyb1}
\rho_*=\frac{1}{\sigma\sqrt{2\pi}}e^{-\frac{x^2}{2\sigma^2}}.
\end{equation}
whose  square root $\rho_*^{1/2}(x)$ $\equiv f(x)$ $=(2\pi \sigma ^2)^{-1/4}$ $\exp (-x^2/4\sigma^2)$
has    Fourier image $(\rho^*)^{1/2}(k)$ $\equiv f(k)$ $=(2\sigma ^2/\pi )^{1/4}$ $\exp (-k^2\sigma ^2)$.
That gives
\begin{equation}\label{dyb4}
{\cal V}_{\mu G}(x)=-\frac{\sigma^{-\mu}}{\sqrt{\pi}}e^{\frac{x^2}{4\sigma^2}}\Gamma\left(\frac{1+\mu}{2}\right)\
_1F_1\left[\frac{1+\mu}{2}, \frac 12, -\frac{x^2}{4\sigma^2}\right],
\end{equation}
where $_1F_1(a,b,x)$ is a hypergeometric function \cite{abr}.
This  $\mu $-family of semigroup potentials   sets solution to the L\'{e}vy stable targeting problem, if  the desired   target has the Gaussian form.

\begin{figure}
\centerline{\includegraphics[width=1.2\columnwidth]{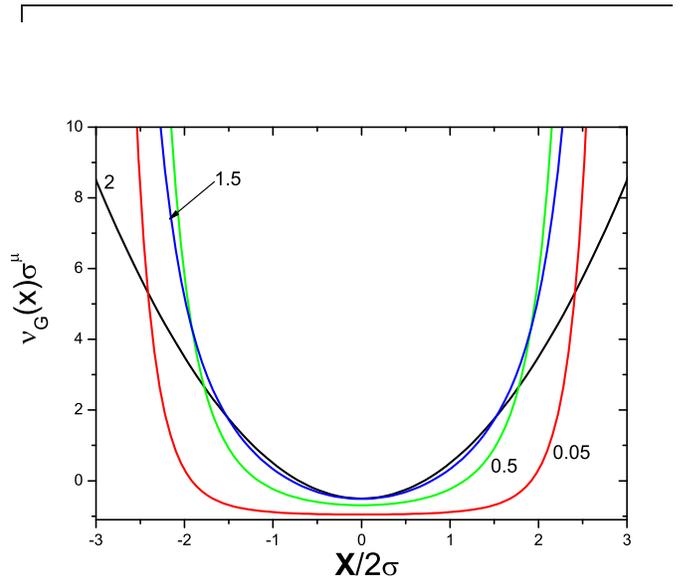}}
\caption{The potential \eqref{dyb4} in normalized variables.
Figures near curves correspond to $\mu$ values.}
\label{f:po}
\end{figure}

Minor comments are necessary  for a qualitative assesment of the  above analytic  result.
The potential ${\cal V}_{\mu G}(x)$ \eqref{dyb4} depends on two parameters: order of fractional derivative $\mu$ and variance $\sigma$.
It can be seen from Eqs. \eqref{dyb1} and \eqref{dyb4} that the  variance $\sigma$ simply alters the width of the potential curve and
does not influence its shape.  The same is true for the factor $\sigma ^{-\mu}$ in front of Eq.\eqref{dyb4}.
That is why in Fig. \ref{f:po} we report the shape of the potential \eqref{dyb4} in normalized variables $z=x/(2\sigma)$
and $y_\mu=\sigma ^{\mu}{\cal V}_{\mu G}(x)$. These universal curves are the same for any $\sigma$ and depend on
 the single parameter $\mu$. Note, that in these variables the $\mu=2$  parabola  assumes the form $y_2=z^2-1/2$.

It is also  seen from Fig. \ref{f:po} that at small $\mu$ the potential $y_\mu$ is around $-1$ (we recollect that at $\mu=0$ ${\cal V}_{\mu G}(x)\equiv -1$), while at larger $x$ it has very steep growth like $\exp(z^2)$. These steep tails flatten as $\mu$ grows and around $\mu=1.5$ the exponential growth of the potential is replaced by power-law $z^\mu$ so that at $\mu=2$ we have the correct asymptotics  $z^2$.

%
\begin{figure}
\centerline{\includegraphics[width=1.1\columnwidth]{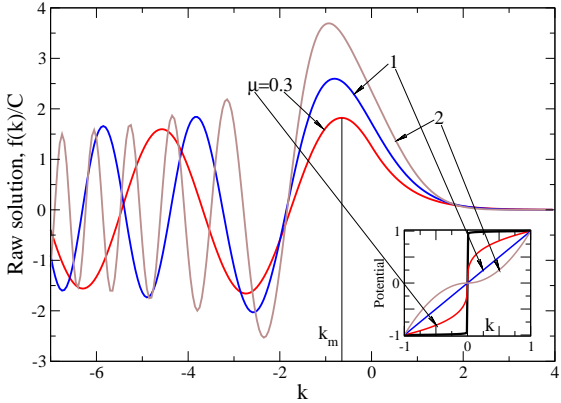}}
\caption{Raw solutions of Eq. \eqref{terp6} (main panel) and potential sign $k \ |k|^\mu$ (inset). Curves are  $\mu$-labeled. Arrows show the correspondence between potential and raw solution for given $\mu$. Thick black line on the inset shows the potential for $\mu=0.01$, which has almost rectangular shape. Solution for $\mu=1$ corresponds to Airy function \eqref{terp6b}.}
\label{f:raw}
\end{figure}

\section{Reverse engineering: asymptotic $\mu$ - targets  for  L\'{e}vy oscillators}

Now we pass to a detailed discussion of  a particular class of  solvable  examples of  the reverse engineering problem  which well illustrates the following general strategy (its full description is moved to  another publication): given a priori  a concrete semigroup with L\'{e}vy driver, infer an asymptotic pdf for the associated master equation (\ref{kinetic}).

Our  main idea  is to  adopt an approach we have developed before, \cite{stef} (see also \cite{lorinczi1,rob}) to the
L\'{e}vy oscillator with ${\cal{V}}(x)=x^2/2$ and arbitrary stability index $\mu$.

We begin with the equation for a terminal pdf $\rho_ *$, inferred from the $\mu $-L\'{e}vy semigroup with a
predefined harmonic  potential
\begin{equation}\label{terp1}
     {\cal V}_\mu(x)\rho ^{1/2}_* \equiv \frac{x^2}{2}\rho ^{1/2}_*= - |\Delta |^{\mu/2}\rho ^{1/2}_*, \ 0<\mu \leq 2.
\end{equation}
We take Fourier images of  both sides of Eq.\eqref{terp1} to obtain
\begin{equation}\label{terp2}
u_k=\frac{1}{\sqrt{2\pi}}\int_{-\infty}^{\infty}\frac{x^2}{2}f(x)e^{\imath kx}dx= -\frac 12 \frac{\partial^2 f(k)}{\partial k^2}.
\end{equation}
The right-hand side of Eq. \eqref{terp1} has the form $-|k|^\mu f(k)$  so that
\begin{equation}\label{terp3}
    \frac{\partial^2 f(k)}{\partial k^2}\equiv \frac{d^2 f(k)}{dk^2} =2|k|^\mu f(k).
\end{equation}

The idea  to solve the Eq. \eqref{terp3} for arbitrary $0<\mu \leq 2$ is borrowed from Ref. \cite{rob}, where the solution for $\mu=1$ had been obtained in terms of Airy functions. The method of Ref. \cite{rob} is based on the consideration of 1D Schr{\"o}dinger problem with a potential being even function of the coordinate, which implies that
the corresponding eigenfunctions should be either even or odd (see e.g. \cite{land3, shiff}). In particular, the ground state wave function should be even as it does not have nodes  \cite{land3}. It can be shown that solution $f(k)$, defining the Fourier image of desired terminal pdf, corresponds to the ground state wave function of the above Schr{\"o}dinger problem. Generalizing the method of Ref. \cite{rob} for arbitrary $\mu$, we can show that to obtain this function for even potential like $|k|^\mu$ we should consider instead of \eqref{terp3} the equation
$\frac{d^2 f(k)}{dk^2} =2\ {\rm{sign}} k \ |k|^\mu f(k)$ or
\begin{equation}\label{subs}
\left\{
\begin{array}{cc}
  \frac{d^2 f(k)}{dk^2} =2k^\mu f(k), & k>0  \\ \\
  \frac{d^2 f(k)}{dk^2} =-2(-k)^\mu f(k), & k<0.   \\
   \end{array}
\right.
\end{equation}

\begin{figure}
\centerline{\includegraphics[width=1.1\columnwidth]{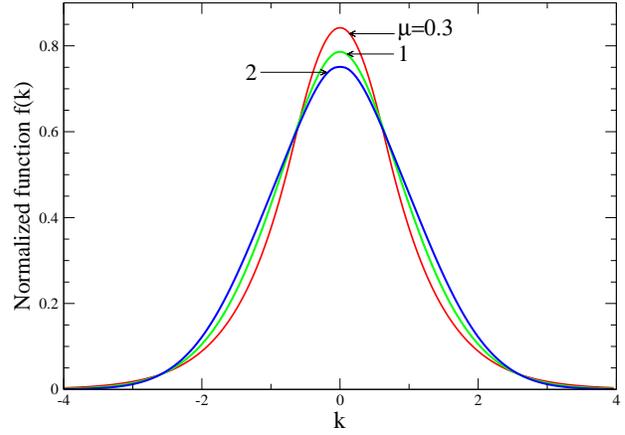}}
\caption{Normalized solutions for Fourier images of square roots of
terminal pdfs in $k$-space. Curves are $\mu$-labeled. } \label{f:normk}
\end{figure}
Now the scenario of obtaining the desired $f(k)$ is as follows. After finding the exponentially decaying solution of Eq. \eqref{subs} for $k>0$ and oscillatory one at $k<0$, we should require the continuity of the function $f(k)$ and its derivative at $k=0$. This is because the Eq. \eqref{subs} is of the second order. After that we should find the position $k_m$ of the first maximum of oscillating part and shift the solution to the right by $k_m$ so that the first maximum of oscillatory part is at $k=0$. Then, "chopping" the rest of oscillating part and reflecting  the obtained piece about the  vertical axis to obtain the even "bell-shaped" function. The resultant  solution in the  $k$ space should be Fourier-inverted and squared to yield  the desired terminal pdf in the $x$ -space.

To fulfill this scenario, we observe the following form of solutions of Eq. \eqref{subs} for  $k>0$ and $k<0$ \cite{polyanin}. Namely, for $k \geq 0$
\begin{equation}\label{terp4}
f(k)=\sqrt{k}\left[C_{11} I_{\frac{1}{2q}}\left(\frac{\sqrt{2}}{q}k^q\right)+
C_{12} K_{\frac{1}{2q}}\left(\frac{\sqrt{2}}{q}k^q\right)\right],
\end{equation}
while for  $k<0$
\begin{equation}\label{terp4m}
f(k)=\sqrt{|k|}\left[C_{21} J_{\frac{1}{2q}}\left(\frac{\sqrt{2}}{q}|k|^q\right)+
C_{22} N_{\frac{1}{2q}}\left(\frac{\sqrt{2}}{q}|k|^q\right)\right],
\end{equation}
where $q=(\mu+2)/2$.
Here $J_\nu(x)$ and $N_\nu(x)$ are Bessel functions and $I_\nu(x)$ and $K_\nu(x)$ are modified Bessel functions, see Ref. \cite{abr}. At $x \to \infty$ $I_\nu(x)$ is exponentially growing function \cite{abr} while $K_\nu(x)$ is exponentially decaying \cite{abr}. On the other hand, as $x \to -\infty$ the functions $J_\nu(x)$ and $N_\nu(x)$ have "needed" oscillatory asymptotics \cite{abr}. This means that to have a localized pdf, we should leave the term with $K_{\frac{1}{2q}}$ in \eqref{terp4} only. Then  $f(k)$ assumes  the  following form
\begin{widetext}
\begin{equation}\label{terp5}
f(k)=\left\{
\begin{array}{cc}
  C_{12}\sqrt{k}K_{\frac{1}{2q}}\left(\frac{\sqrt{2}}{q}k^q\right), & k \geq 0  \\ \\
  \sqrt{|k|}\left[C_{21} J_{\frac{1}{2q}}\left(\frac{\sqrt{2}}{q}|k|^q\right)+
C_{22} N_{\frac{1}{2q}}\left(\frac{\sqrt{2}}{q}|k|^q\right)\right], & k<0.   \\
   \end{array}
\right.
\end{equation}
\end{widetext}
Now we join (glue) the obtained solutions at $k=0$ to secure a continuity of a function and its first derivative.

The gluing  procedure yields
\begin{equation}\label{terp6}
f(k)=C\sqrt{|k|}\left\{
\begin{array}{cc}
  K_\nu(u), & k \geq 0  \\ \\
  \frac{\pi}{2}\left[\cot \frac{\pi \nu}{2} J_\nu(u)-
 N_\nu (u)\right], & k<0,  \\
   \end{array}
\right.
\end{equation}
where $C \equiv C_{12}$,
\begin{equation}\label{terp6a}
\nu=\frac{1}{2q}\equiv \frac{1}{\mu+2},\ u=\frac{\sqrt{2}}{q}|k|^q\equiv \frac{2\sqrt{2}}{\mu+2}|k|^{1+\frac{\mu}{2}}.
\end{equation}
We note here that for  the Cauchy driver, i.e.  $\mu=1$ we obtain from \eqref{terp6} the result
\begin{equation}\label{terp6b}
f(k)=C\sqrt{k}K_{\frac{1}{3}}\left(\frac{2\sqrt{2}}{3}k^{\frac 32}\right)=
C\frac{\pi \sqrt{3}}{2^{\frac 16}}{\rm {Ai}}\left(2^{\frac 13}\ k\right),
\end{equation}
known from  our earlier publication \cite{stef}.

The "raw" solutions \eqref{terp6} are plotted on the main panel of Fig. \ref{f:raw} for different values of $\mu$. It is seen from the inset that for $\mu \to 0$ (thick black line corresponding to $\mu$=0.01) the potential has the shape of almost rectangular barrier, corresponding to decaying solution (localized particle inside the barrier) at $k>0$ and oscillating one (free particle) at $k<0$ \cite{land3, shiff}. We note here that for potentials depicted on the inset to Fig.
\ref{f:raw} the above kind of solution exist only if its eigenenergy lies between the limiting values of a barrier at $|x| \to \infty$  \cite{land3, shiff}. In this case the zeroth eigenenergy, which is the case for Eqs. \eqref{terp3} and \eqref{subs} perfectly suits the problem under consideration not only for $\mu \to 0$, where the barrier is almost rectangular, but also at higher $\mu$. This explaines the fact that as the shape of barrier deviates from rectangular one at $\mu$ increase, the oscillations at $k<0$ start to decay, the strongest one being at $\mu=2$. Also, with the growth of $\mu$, the period of the oscillations lowers, the  minimum being achieved  at $\mu=2$ also.

\begin{figure*}
\centerline{\includegraphics[width=1.1\columnwidth]{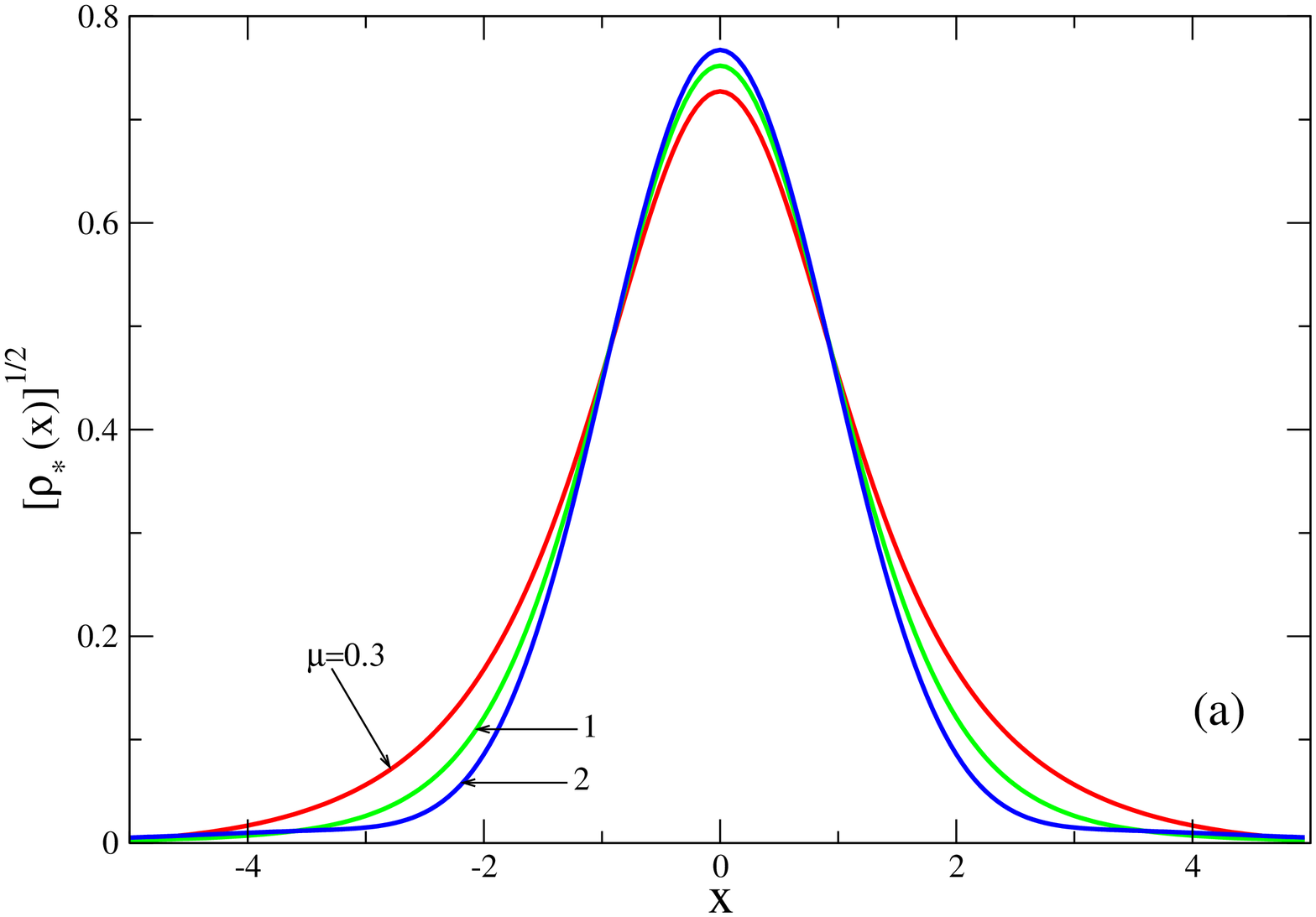}
\includegraphics[width=1.1\columnwidth]{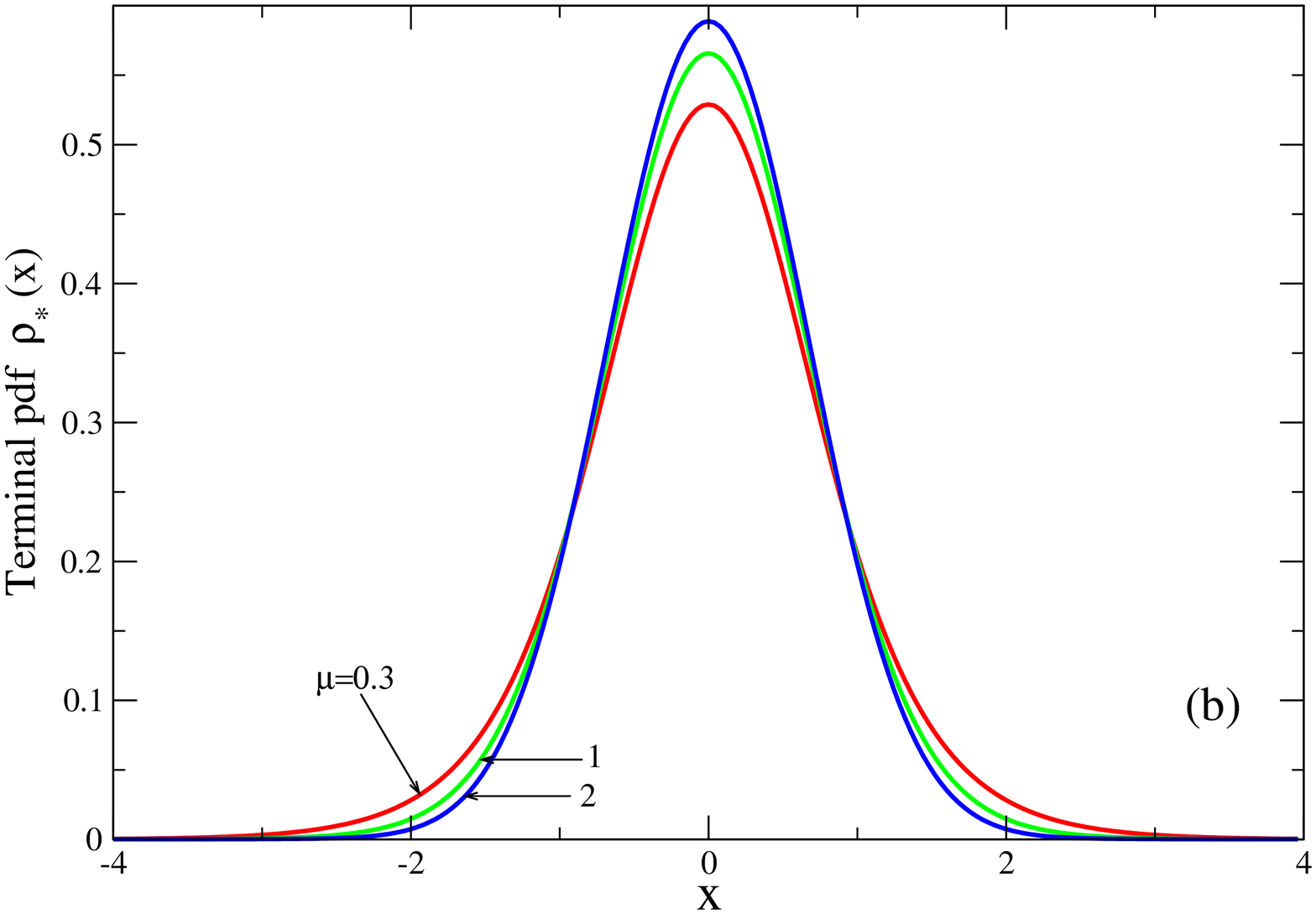}}
\centerline{\includegraphics[width=1.1\columnwidth]{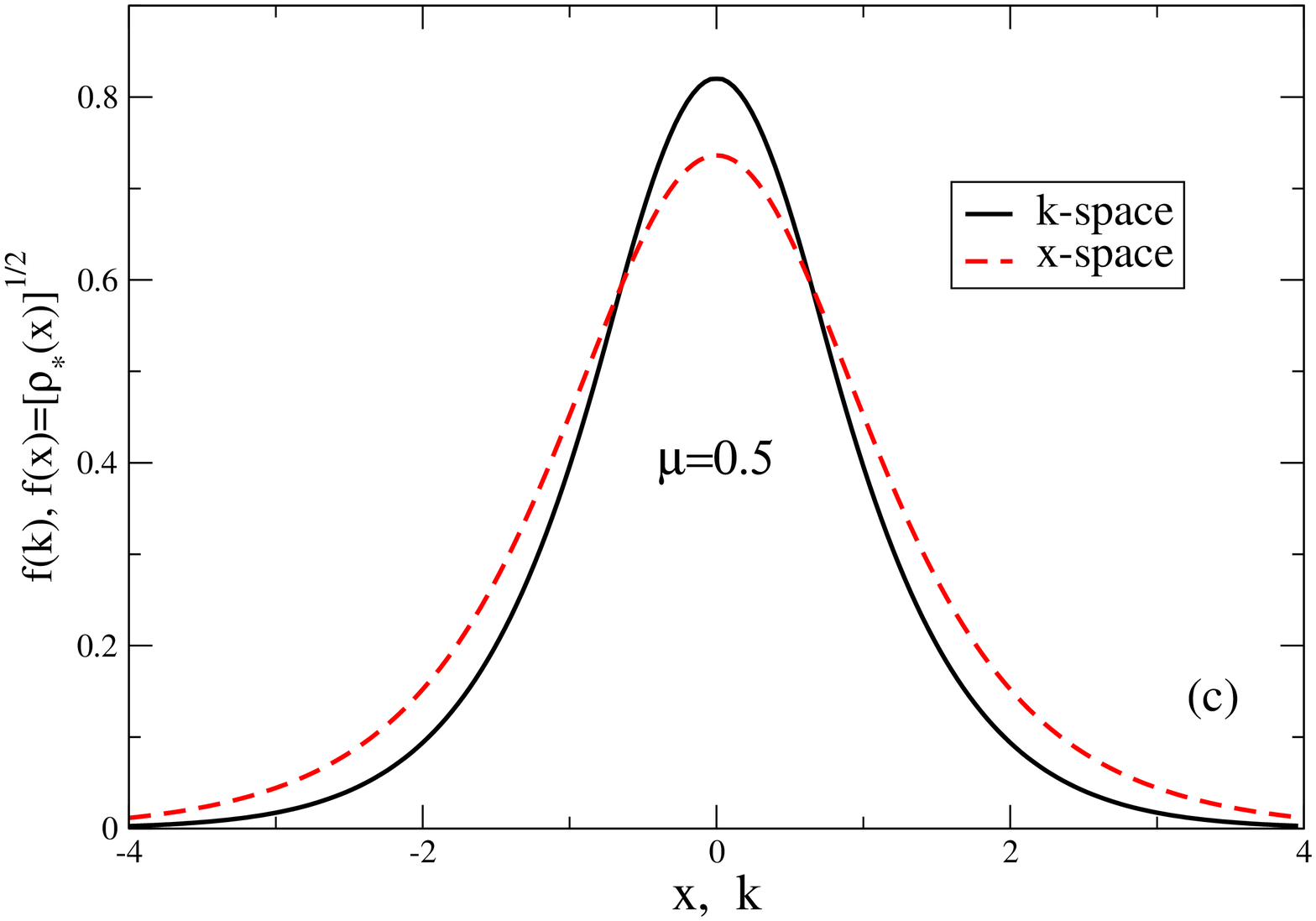} }
\caption{Panel (a) - inverted Fourier images $[\rho_*(x)]^{1/2}$;
panel (b) - desired terminal pdfs at different $\mu$ (figures).
Panel (c) compares the  behavior of the functions in $k$ and
$x$-spaces for $\mu=0.5$.} \label{f:xprvo}
\end{figure*}

Now we find the position $k_m$ of the first maximum of oscillating part. Equating to zero the first derivative of an  oscillating part of \eqref{terp6} we arrive at
\begin{equation}\label{max1}
N_{\nu-1}(u)-\cot\frac{\pi \nu}{2}J_{\nu-1}(u)=0,
\end{equation}
where $\nu$ and $u$ are defined by \eqref{terp6a}.  The roots of Eq.~\eqref{max1} can easily  be obtained numerically for different $\mu $.

The normalization of the obtained function can be achieved through the condition $C^2\int_{-\infty}^{\infty}f^2(k)dk=1$ or
\begin{equation}\label{nomr}
2C^2\left[\int_{0}^{-k_m}f_1^2(k)dk+
\int_{-k_m}^{\infty}f_2^2(k)dk\right]=1,
\end{equation}
where $f_1$ and $f_2$ denote  oscillatory and decaying parts of Eq. \eqref{terp6} respectively.
 Normalized solutions in the  $k$-space for different $\mu$'s are reported in Fig. \ref{f:normk}. It is seen that for  small $k$ and on the tails,
  the distribution functions for higher $\mu$'s run below those for smaller $\mu$'s, while in the intermediate $k$ range the situation is opposite.

The final step of the procedure is to invert the $k$-space solutions to  the  $x$-space and square them to obtain the desired terminal pdf. For general $\mu$ this procedure can be accomplished only numerically.

Fig. \ref{f:xprvo}  displays  both the inverted functions $f(k)$, corresponding to square roots of the inferred  terminal pdfs (panel (a)) and those pdfs themselves (panel (b)). The opposite (if compared to this  in the  $k$-space) tendency is seen  in the  $x$ space, where the curve corresponding to lowest $\mu$ lies below all other curves  in the  small $x$ region and has slowest decay. As $\mu$ grows, the central part of the curve rises and tails become steeper.

Panel (c) of Fig. \ref{f:xprvo} reports a comparison between the shapes of functions $f(k)$ and $f(x)$. The situation here is the same as that for  the  Airy function, as discussed in \cite{gar1}. Namely, the function in $k$-space decays quicker then in $x$-space and its value at the center is larger then that in $x$ - space.
We plot here the exemplary  case of $\mu=0.5$, the situation for other $\mu$ is qualitatively the same.

\section{Outlook}

The next natural step in our $\mu$-targeting procedure is to obtain (numerically) the dynamics of a function $\rho(x,t)$ for
L\'evy oscillators with different values of  $\mu$. This can be done both for  the semigroup process \eqref{mutar2} and for  the
 Langevin-driven one  (e.g. fractional Fokker-Planck dynamics).  Those patterns of temporal behavior are   inequivalent, although
 both processes  may terminate at common  pdfs with a predefined decay at infinities.  The latter  pdfs  may have heavy tails, but generically
 admit an  arbitrary (finite, eventually infinite)  number of moments.

A more general  problem  would be that  of the   existence of terminal pdfs, after passing from the master equation to the  (fractional)
 Hamiltonian dynamics   \eqref{hamiltonian} with  an  arbitrary
potential ${\cal V}$,  in one, two or three spatial dimensions.

We note that in the case of
$\mu=2$, when fractional Hamiltonian \eqref{hamiltonian} reduces to ordinary quantum-mechanical Hamiltonian operator.
In  the standard  quantum mechanical  setting  (see, e.g., Refs. \cite{land3, shiff}) the above question is equivalent
to an issue  of the existence of  bound states in a given potential.  The  quantum
 mechanical language appears because  we can convert the parabolic equation of the  Fokker-Planck type to
the generalized Schr\"{o}dinger equation.

 The wave function of a  bound state
should be  localized  to ensure  a normalization of its squared expression, i.e. the  corresponding stationary  pdf of the Fokker-Planck equation.
It is known  (see, e.g., Ref. \cite{land3}) that in 1D case the bound state exist in the potential well $U(x)$ of not only finite but an
 infinitesimal depth.
  The only restriction is that the integral $\int_{-\infty}^\infty U(x)dx$ should exist. The  latter condition is equivalent to the
   requirement that $U(x)$ should have the same asymptotics at infinities and potential zero point $U(\pm \infty)=0$.
    In the 2D case, when the potential $U=U(x,y)$, the situation is similar to that in 1D one, while in 3D ($U=U(x,y,z)$) the situation is
    to some extent opposite - if the potential well is not sufficiently deep (see Ref.\cite{land3} for details), the particle cannot
     be "captured", so that bound state does not exist.  Confining potentials  in 3D, where bound states exist, form the  so-called Kato class
     of potentials.

 The presence of fractional derivatives with $0<\mu \leq 2$ alters the picture both in 1D (2D) and in 3D. In 1D they  definitely  "spoil" the
 bound states. It is not only that the pdfs  (if in existence)  may  have  heavier tails  if  compared to the   conventional  ($\mu =2$) case.
 The pdfs in question may not exist at all, if  a normalizability of the bound state is lost.
 In 3D and in equations with fractional derivatives there  may typically  be  no  normalizable   bound states (and thus terminal pdfs),
  except for a carefully selected (Kato)-subclass  of conceivable potentials.

Some peculiarities pertaining to the (non)-existence of invariant pdfs in the case of L\'{e}vy   drivers  (Langevin-driven fractional dynamics)
 were discussed for 1D case in Ref.~\cite{dybiec}.
We have encountered the same problem in connection with  the Cauchy family of pdfs \cite{stef,stef1}, see also Ref. \cite{barkai} for
a discussion of so-called infinite covariant densities.

\end{document}